\documentclass[%
 prl,%
 amsmath,%
 amssymb,%
 twocolumn,%
 superscriptaddress,%
  reprint,
]{revtex4-1}

\usepackage{graphicx}
\usepackage{subfigure}
\usepackage{dcolumn}
\usepackage{bm}
\usepackage{ulem}
\usepackage{color}

\begin{document}


\title{Why graphene growth is very different on the C face than on the Si face of SiC: Insights from surface equilibria and the (3$\times$3)-3C-SiC($\bar{\text{1}}\bar{\text{1}}\bar{\text{1}}$) reconstruction}

\author{Lydia Nemec}
\affiliation{Fritz-Haber-Institut der Max-Planck-Gesellschaft, D-14195, Berlin, Germany}
\author{Florian Lazarevic}
\altaffiliation{Present address: AQcomputare GmbH, Business Unit MATcalc, Annabergerstr. 240, 09125 Chemnitz, Germany}
\author{Patrick Rinke}
\affiliation{Fritz-Haber-Institut der Max-Planck-Gesellschaft, D-14195, Berlin, Germany}
\affiliation{COMP/Department of Applied Physics, Aalto University, P.O. Box 11100, Aalto FI-00076, Finland}
\author{Matthias Scheffler}
\affiliation{Fritz-Haber-Institut der Max-Planck-Gesellschaft, D-14195, Berlin, Germany}
\author{Volker Blum}
\affiliation{Fritz-Haber-Institut der Max-Planck-Gesellschaft, D-14195, Berlin, Germany}
\affiliation{Department of Mechanical Engineering and Material Science, Duke University, Durham, NC 27708, USA}

\date{\today}

\begin{abstract}
We address the stability of the surface phases that occur on the C-side of 3C-SiC($\bar{1} \bar{1} \bar{1}$) at the onset of graphene formation. In this growth range, experimental reports reveal a coexistence of several surface phases. This coexistence can be explained by a Si-rich model for the unknown (3$\times$3) reconstruction, the known (2$\times$2)$_{C}$ adatom phase, and the graphene covered (2$\times$2)$_{C}$ phase. By constructing an \textit{ab initio} surface phase diagram using a van der Waals corrected density functional, we show that the formation of a well defined interface structure like the ``buffer-layer'' on the Si side is blocked by Si-rich surface reconstructions.
\end{abstract}


\maketitle

Graphene grown on silicon carbide (SiC) is one of the most promising material combinations for future graphene applications.\cite{berger2006eca, emtsev2009tws, lin2011wsg, her2012ttg, deheer2011eg, guo2013rmo}
On SiC, graphene growth is achieved by thermal decomposition of the substrate.\cite{bommel1975laa, forbeaux2000ssg, berger2004ueg} The electronic properties of few-layer graphene films grown on the C-side of the polar SiC surface are similar to those of an isolated monolayer graphene film with very high electron mobilities.\cite{hass2008wmg, sprinkle2009fdo, berger2010epo} However, controlling the layer
thickness of the graphene films remains a challenge.\cite{mathieu2011mcb} While some groups report the successful growth of large-scale monolayer graphene \cite{hu2012seg, ruan2012ego}, other reports suggest that the pure monolayer growth regime is difficult to achieve on the C face of SiC. \cite{mathieu2011mcb} This is very different from the Si side, where nearly perfect, monolayer graphene films can be grown over large areas.\cite{emtsev2009tws, deheer2011eg} Investigating the relative phase stability of the competing surface phases in the thermodynamic range of graphitisation is an important step for a better understanding of graphene growth.

For epitaxial graphene films on the Si-side of 3C-SiC(111), we have recently shown by \textit{ab initio} atomistic thermodynamics that individual phases, the ($6 \sqrt{3} \times 6 \sqrt{3}$)-R30$^{\circ}$ zero layer and monolayer graphene (ZLG and MLG) can form as near equilibrium phases under certain external conditions (represented by the C and Si chemical potentials that are controlled by temperature and background gas pressure in experiment, see also Eq. (\ref{Eq:Esurf}) below).\cite{nemec2013tec} What is not \textit{a priori} clear is whether on the C face of SiC, graphene films can also be thermodynamically stable. To address this question, atomistic models are required for the different competing phases, in particular for the C rich conditions close to the graphitization regime.

We here present first-principles evidence that the formation of monolayer graphene films on the C face is hindered by stable Si rich phases. This is a major and unexpected difference to the case of the Si face, where the formation of a C-rich, so-called ``buffer layer'' phase is actually aided by the formation of heterogeneous C-Si bonds. To shed light on the phase mixture at the graphitisation limit on the C face, we use a possible model of the unknown (3$\times$3) reconstruction. The central feature of this model is a capping layer of Si atoms, minimizing dangling bonds in the same way as the known Si-rich (3$\times$3) phase on the Si face of SiC.\cite{starke1998nrm, schardt2000cot}

On the C-side, a series of different surface structures have been observed during annealing.\cite{bommel1975laa, starke1997aso, hoster1997maa, li1996aso, bernhardt1999ssr, seubert2000iss, magaud2009got, hiebel2009aae, starke2009ego, hiebel2012sas} Graphene growth starts either with a Si rich (2$\times$2) phase in an Si rich environment\cite{bernhardt1999ssr, li1996aso} or with an oxidic ($\sqrt{3} \times \sqrt{3}$) reconstruction.\cite{bernhardt1999eio,  li1996aso} In the absence of a Si background like disilane, a (1$\times$1) phase is observed, which exhibits the periodicity of bulk SiC underneath a disordered oxidic layer.\cite{starke1997aso} Continued heating leads to a (3$\times$3) phase. Using qualitative low-energy electron diffraction (LEED), Bernhardt \textit{et al.} \cite{bernhardt1999ssr} showed that the (3$\times$3) reconstructions originating from different starting structures and environments are equivalent.\cite{bernhardt1999ssr} Further annealing leads to a (2$\times$2) Si ad-atom phase, referred to as (2$\times$2)$_\text{C}$ (notation taken from \cite{bernhardt1999ssr}). Just before graphene forms on the surface, a coexistence of the two surface phases, the (3$\times$3) as well as the (2$\times$2)$_\text{C}$, is observed. \cite{bernhardt1999ssr, veuillen2010iso} In addition, different groups reported strong experimental evidence that both reconstructions persist underneath the graphene films with the (3$\times$3) phase gradually fading, but never disappearing.\cite{hass2007spo, hiebel2008gsi, emtsev2008iga, hass2008tga, starke2009ego, veuillen2010iso, hiebel2012sas, moreau2013hra} While the atomic structure of the (2$\times$2)$_\text{C}$ reconstruction was resolved by quantitative LEED \cite{seubert2000iss}, the (3$\times$3) reconstruction and the graphene/ SiC interface remain a puzzle.
\begin{figure}
  \includegraphics[width=0.5\textwidth]{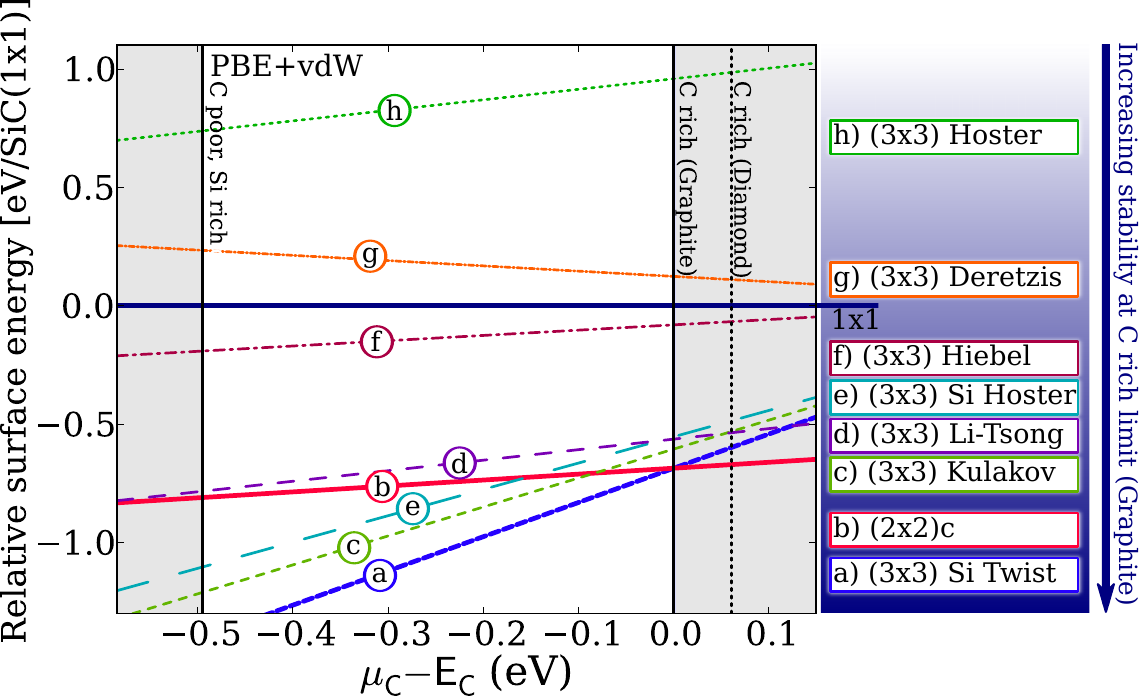}
  \caption{Comparison of the surface energies relative to the bulk terminated (1$\times$1) phase as a function of the C chemical potential within the allowed ranges (given by diamond Si and graphite C). Shaded areas indicate chemical potential values outside the strict thermodynamic stability limits. The surface energy diagram includes structure models proposed earlier for the C face [(b)\cite{seubert2000iss} (d)\cite{li1996aso}, (e) and (h)\cite{hoster1997maa}, (f)\cite{hiebel2012sas}, (g)\cite{deretzis2013adf}] and models adapted from the Si face [(a)\cite{starke1998nrm} and (c)\cite{kulakov1996ssr}]. }
\label{Fig:3x3_Surface_energies}
\end{figure}
In the past, several structural models were suggested for the (3$\times$3) phase on the C face.\cite{li1996aso, hoster1997maa, hiebel2012sas, deretzis2013adf} We here summarise its experimentally observed characteristics. Its stoichiometry was found to be Si-rich by Auger electron spectroscopy (AES).\cite{hoster1997maa, bernhardt1999ssr} The corresponding filled state scanning tunneling microscope (STM) image is consistent with three adatoms residing at the same height.\cite{deretzis2013adf, hiebel2012sas} Scanning tunnelling spectroscopy shows a semiconducting surface with a $1.5$~eV band gap.\cite{hiebel2009aae}

For the graphene/SiC interfaces on the C-side two different scenarios have been invoked. a) The first carbon layer is strongly bound to the substrate.\cite{varchon2007eso, mathieu2011mcb, srivastava2012iso} In this scenario, the Si sublimation rate during graphene growth is controlled by either working in an inert gas atmosphere \cite{emtsev2009tws}, by using a confined geometry \cite{hass2008tga}, or by providing an external Si gas phase \cite{srivastava2012iso, forbeaux2000ssg} for example disilane. It is not clear, however, if the different groups observe the same structures. Based on their LEED data, Srivastava \textit{et al.} \cite{srivastava2012iso} proposed a $\left( \sqrt{43} \times \sqrt{43} \right)$-R$\pm 7.6 ^{\circ}$ SiC substrate with a (8$\times$8) carbon mesh rotated by $7.6 ^{\circ}$ with respect to the substrate ('$\sqrt{43}$-R$7.6 ^{\circ}$').\cite{srivastava2012iso} The detailed atomic structure of this interface is not known.\cite{guowei2013foa}  b) For samples prepared under ultra-high vacuum conditions, the first carbon layer is weakly bound to the substrate, showing the characteristic behavior of the $\pi$-band at the $K$-point of the Brillouin zone. Here an inhomogeneous interface is present since the (3$\times$3) as well as a (2$\times$2)$_{\text{C}}$ reconstruction is observed underneath the graphene layers.\cite{hiebel2008gsi, emtsev2008iga, starke2009ego} A recent study on graphene grown by molecular beam epitaxy (MBE) on the C-side exhibited the same structural characteristics as graphene grown by high-temperature annealing.\cite{moreau2013hra} This is a strong indication that indeed the (2$\times$2)$_{\text{C}}$ as well as the (3$\times$3) reconstruction prevails below the graphene films.

To our knowledge, the different (3$\times$3) models suggested in the literature and likewise the remaining phases including the graphene/SiC interfaces have not yet been placed in the context of a surface phase diagram. We employ density-functional theory (DFT) using the FHI-aims all-electron code \cite{blum2009aim,havu2009eoi} with the ELPA eigensolver library.\cite{auckenthaler2011pso, marek2014ELPA} We use the van der Waals (vdW) corrected \cite{tkatchenko2009amv}  Perdew-Burke-Enzerhof (PBE) generalised gradient approximation\cite{perdew1997gga} (PBE+vdW) and the Heyd-Scuseria-Ernzerhof hybrid functional (HSE06+vdW) \cite{krukau2006iot, tkatchenko2009amv} for the exchange correlation functional. Unless otherwise noted the calculations are non-spinpolarised. Technical parameters and bulk lattice constants are listed in the supplemental material (SM)\cite{supplementary}.  In contrast to hexagonal polytypes, at the surface of 3C-SiC only one type of stacking order is present \cite{starke1997aso}. Although the growth process \cite{lavia2014mog} and the electronic structure \cite{pankratov2012goc} differ among polytypes, the surface reconstruction does not seem to be affected \cite{furthmuller1998mot, schardt2000cot, pankratov2012goc}.

A good indicator for finding the most likely (3$\times$3) reconstruction or interface structures is a comparison of the respective surface free energies as formulated in the \textit{ab initio} atomistic thermodynamics approach.\cite{weinert1986cav, scheffler1988tao, scheffler1988pfc, reuter2002csa, reuter2005book} We neglect vibrational and configurational entropy contributions to the free energy, although in the coexistence region they might lead to small shifts. In the limit of sufficiently thick slabs, the surface energy $\gamma$ of a two-dimensional periodic SiC slab with a C face and a Si face is given as
\begin{equation}\label{Eq:Esurf}
  \gamma_\text{Si face} + \gamma_\text{C face} = \frac{1}{A} \left(
  E^\text{slab} - N_\text{Si} \mu_\text{Si} -
  N_\text{C} \mu_\text{C} \right) \, .
\end{equation}
$N_\text{Si}$ and $N_\text{C}$ denote the number of Si and C atoms in the slab, respectively.  $\mu_\text{Si}$ and $\mu_\text{C}$  refer to the chemical potentials of Si and C. The stability of the SiC bulk dictates $\mu_\text{Si} + \mu_\text{C} = E^\text{bulk}_\text{SiC}$ where $E^\text{bulk}_\text{SiC}$ is the total energy of a bulk SiC unit cell. All surface energies are given in eV per area ($A$) of a (1$\times$1) SiC unit cell. The letter $E$ denotes total energies for a given atomic geometry throughout this work.
The chemical potential limits of the C and Si reservoirs are fixed by the requirement that the underlying SiC bulk is stable against decomposition~\cite{nemec2013tec}, leading to: $E^\text{bulk}_\text{SiC} - E^\text{bulk}_\text{Si} \le \mu_\text{C} \le E^\text{bulk}_\text{C}$. Because of the close competition between the diamond and graphite structure for C \cite{berman1955,yin1984, choljun2014rpc}, we include both limiting phases in our analysis.

The surface energies of the (2$\times$2)$_\text{C}$ surface model by Seubert \textit{et al.} and the different models for the SiC-(3$\times$3) reconstruction are shown as a function of $\Delta \mu_\text{C} = \mu_\text{C}-E^\text{bulk}_\text{C}$ in Fig.~\ref{Fig:3x3_Surface_energies}. The structure with the lowest energy for a given $\Delta \mu_\text{C}$  corresponds to the most stable phase. We here briefly discuss the alternative surface phases of the SiC C face  (for more details see SM \cite{supplementary}). Hoster, Kulakov, and Bullemer suggested a geometric configuration for the (3$\times$3) reconstruction on the basis of STM measurements without specifying the chemical composition.\cite{hoster1997maa} Highest in surface energy is the variation suggested by Hiebel \textit{et al.} \cite{hiebel2012sas} labeled \textit{h}, followed by a carbon rich composition suggested by Deretzis and La Magna \cite{deretzis2013adf} labeled \textit{g}. We added a modification with all adatoms chosen to be Si, labeled \textit{e}. Of all the Hoster-type models that we tested, this is the most stable chemical composition. Hiebel \textit{et al.} suggested a new model, labeled \textit{f}. \cite{hiebel2012sas} Li and Tsong proposed a tetrahedrally shaped cluster as reconstruction.\cite{li1996aso} They suggested a Si and C rich configuration. We tested both configurations and included an additional Si tetrahedron (for details see SM \cite{supplementary}). Here, we include only the most stable cluster formed by 4 Si atoms, labeled \textit{d}. Finally, we added a model originally proposed as a Si rich structure for the 6H-SiC($0001$)-(3$\times$3) reconstruction by  Kulakov, Henn, and Bullemer \cite{kulakov1996ssr}, labeled \textit{c}. To summarise Fig.~\ref{Fig:3x3_Surface_energies}, all the alternative models we tested are too high in energy at the graphite line to coexist with the (2$\times$2)$_\text{C}$ adatom model. 
\begin{figure}
  \includegraphics[width=0.40\textwidth]{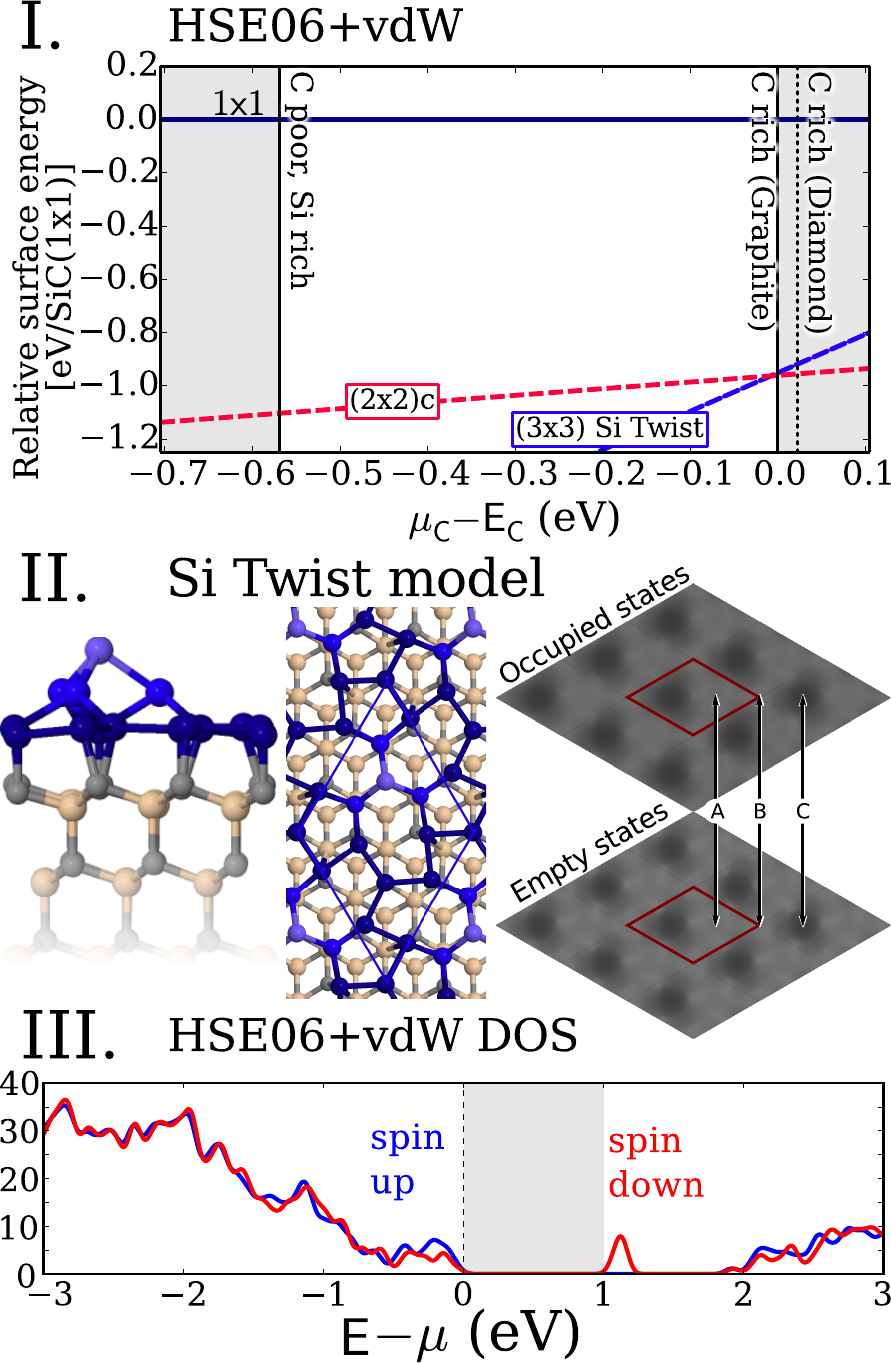}
  \caption{\label{Fig:3x3_Surface_energies_HSE} I: The a) and b) phases from Fig.~\ref{Fig:3x3_Surface_energies} calculated using the HSE06 exchange-correlation functional with fully relaxed structures and unit cells. II: The geometry of the \textit{Si twist model} and simulated constant current STM images for occupied and empty state of the \textit{Si twist model} (unit cell shown in red). The three points of interest (A, B, C) marked by arrows are labeled according to Hiebel \textit{et al.}\cite{hiebel2012sas}.
  III: Spin-polarized density of states.}
\end{figure}

We next show that a conceptual model for the (3$\times$3) reconstruction based on a Si-rich termination performs much better at explaining the various surface characteristics. To create a plausible termination, we base our model on the \textit{Si twist model} \cite{starke1998nrm, schardt2000cot}, known from the 3C-SiC($111$)--(3$\times$3) reconstruction. In Fig.~\ref{Fig:3x3_Surface_energies_HSE}~(II) its geometry is shown in a side view and from atop. The top bulk C layer is covered by a Si adlayer forming heterogeneous Si-C bonds. Three Si adatoms form a triangle twisted by $7.7^{\circ}$ with respect to the top SiC layer. In comparison, the twist angle on the Si side amounts to $9.3^\circ$. The topmost Si adatom is positioned on top of the triangle. In the surface diagram (Fig.~\ref{Fig:3x3_Surface_energies}), this phase has the lowest energy of all previously proposed (3$\times$3) models. Its formation energy crosses that of the (2$\times$2)$_\text{C}$ phase just at the C rich limit (graphite) of the chemical potential. To coexist with the (2$\times$2)$_\text{C}$ phase and to be present at the onset of graphite formation,  the (3$\times$3) phase has to cross the graphite line very close to the crossing point between the graphite line and the (2$\times$2)$_\text{C}$ phase. The \textit{Si twist model} shown in Fig.~\ref{Fig:3x3_Surface_energies_HSE} satisfies this condition.

In agreement with the AES experiments our model is Si-rich. We also compared the surface energetics of the coexisting phases (2$\times$2)$_\text{C}$  and the \textit{Si twist model} using the higher level HSE06+vdW hybrid functional with fully relaxed structures and unit cells, shown in Fig.~\ref{Fig:3x3_Surface_energies_HSE}~(I). As can be seen, the phase coexistence does not depend on the chosen functional. However, to distinguish between a phase coexistence or a close competition between the two phases, the inclusion of entropy terms would be the next step.\cite{freibelman2004eoh} 

In Fig.~\ref{Fig:3x3_Surface_energies_HSE}~(III), we show the spin-polarised electronic density of states (DOS) for the spin down (in red) and spin up (in blue) channel. While the spin down surface state gives rise to a peak in the band gap above the Fermi-level, the spin up surface state is in resonance with the SiC bulk states. The DOS clearly demonstrates that the surface is semiconducting in agreement with experiment, featuring a band gap of $1.12$~eV. Thus, the conceptual \textit{Si twist model} - inspired by the Si side - appears to satisfy the existing experimental constraints well.

Furthermore, our simulated STM images Fig.~\ref{Fig:3x3_Surface_energies_HSE}~(II) reproduce the measured height modulation,\cite{hiebel2012sas} but the experimentally observed difference in intensity between occupied and empty state images is not captured by our simulated images. The disagreement in the STM images might be an indication that a different structure is observed in the STM measurements. We started an exhaustive structure search \cite{jan2014} to find a surface model that is even lower in energy than the \textit{Si twist model} and that reproduces all experimental observations, including the STM images. However, if an alternative model were to be found, its surface energy would have to be close to the \textit{Si twist model} at the graphite line to still coexist with the (2$\times$2)$_\text{C}$ phase. As a result, it would very likely coexist with the \textit{Si twist model}.

\begin{figure}
  \includegraphics[width=0.5\textwidth]{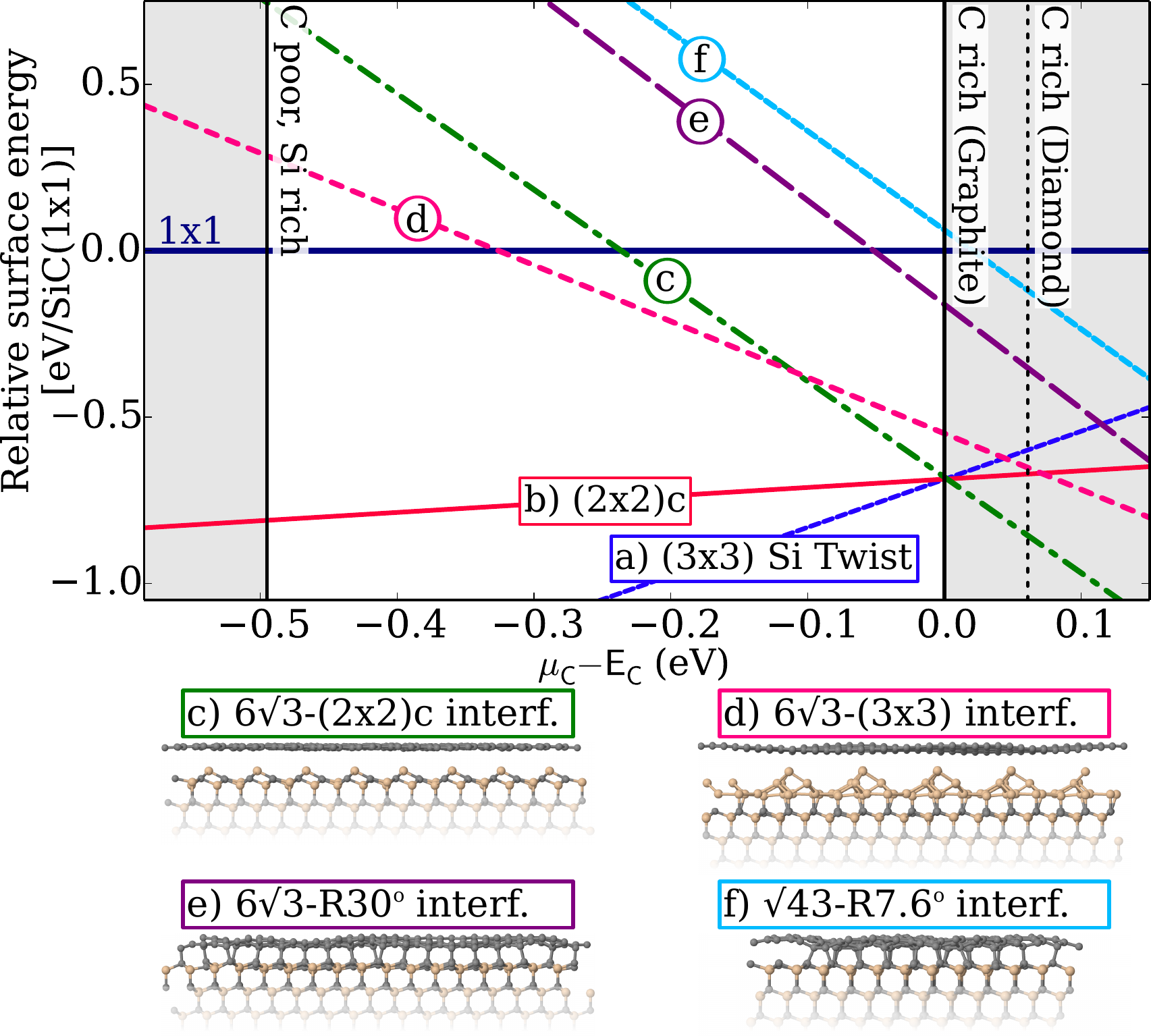}
  \caption{ Comparison of the surface energies for four different interface structures of 3C-SiC($\bar{1} \bar{1} \bar{1}$), relative to the bulk-truncated (1$\times$1) phase, as a function of the C chemical potential within the allowed ranges (given by diamond Si, diamond C or graphite C, respectively),  using the graphite limit as zero reference. In addition, the known (2$\times$2)$_\text{C}$ reconstruction and the (3$\times$3) \textit{Si twist model} are shown.}
\label{Fig:Interfaces}
\end{figure}

In the following we use the \textit{Si twist model} as a representative model to shed light on the SiC-graphene interface on the C face. In particular, we will make a simple qualitative argument why the C-rich ZLG interface, critical towards MLG formation on the Si face, does not form. Figure~\ref{Fig:Interfaces} shows the PBE+vdW surface energies of four different interface structures, the (2$\times$2)$_\text{C}$ surface phase and the proposed (3$\times$3) \textit{Si twist model}.

As a first step we constructed an interface structure similar to the ZLG phase known from the Si face - a 6 $\sqrt{3}$-R30$^{\circ}$-interface, labeled \textit{e} in Fig.~\ref{Fig:Interfaces}. This structure crosses the graphite line $0.46$~eV above the crossing point of the (3$\times$3) \textit{Si twist model}, rendering it unstable. As a second structure, we included a purely C based model of the $\sqrt{43}$-R 7.6$^\circ$-interface, labeled \textit{f} in the surface phase diagram of Fig.~\ref{Fig:Interfaces}. In our calculations, this structure is even higher in energy than the $6 \sqrt{3}$-R30$^{\circ}$-interface.

The 6 $\sqrt{3}$-R30$^{\circ}$- and $\sqrt{43}$-R 7.6$^\circ$-interface are models of a strongly bound interface. Since the Gibbs energy of formation for SiC is quite large ($-0.77$~eV in experiment \cite{kleykamp1998geo}, $-0.56$~eV in DFT-PBE+vdW and $-0.59$~eV in DFT-HSE06+vdW), the formation of SiC bonds is favorable. This explains why the (2$\times$2)$_\text{C}$ and (3$\times$3) \textit{Si twist model} are more stable, because they contain a large number of Si-C bonds. Conversely, the $6 \sqrt{3}$-R30$^{\circ}$- and the $\sqrt{43}$-R 7.6$^\circ$-interface are made up of energetically less favorable C-C bonds, which increases the surface energy considerably.

For the weakly bound interface a (2$\times$2) and (3$\times$3) LEED pattern was observed underneath graphene \cite{hiebel2008gsi, emtsev2008iga, starke2009ego}. A typical feature of the LEED structure is a ring like pattern originating from the rotational disorder of graphene films grown on the C face.\cite{hass2006hog, hass2008wmg, hiebel2008gsi, veuillen2010iso, hiebel2008gsi} To model the interface, we limited our study to a 30$^\circ$ rotation between the substrate and the graphene film. This choice was motivated by the LEED study of Hass \textit{et al.}\cite{hass2008wmg}, who showed that graphene sheets on the C face appear mainly with a 30$^\circ$ and a $\pm 2.2^\circ$ rotation. A 30$^\circ$ rotation has also be seen in STM measurements for the graphene covered (2$\times$2) and (3$\times$3) phases \cite{hiebel2008gsi}, from here on called (2$\times$2)$_{\text{G}}$ and (3$\times$3)$_{\text{G}}$. We therefore chose a ($6 \sqrt{3} \times 6 \sqrt{3}$) SiC supercell covered by a $\left( 13\times 13 \right)$ graphene cell rotated by 30$^{\circ}$ with respect to the substrate.

The (2$\times$2)$_{\text{G}}$-interface covers 27 unit cells of the (2$\times$2)$_\text{C}$ reconstruction (labeled \textit{c} in Fig.~\ref{Fig:Interfaces}). It crosses the (2$\times$2)$_\text{C}$ reconstruction just to the right of the graphite limit at a chemical potential of $2$~meV and a surface energy of $-0.69$~eV. This finding demonstrates that the observed $\left( 2 \times 2 \right)$ LEED pattern underneath the graphene layer is indeed consistent with the well known (2$\times$2)$_\text{C}$ reconstruction. Our model of the graphene covered (3$\times$3) phase consist of the same (13$\times$13) graphene supercell, covering 12 units of the (3$\times$3) \textit{Si twist model} (labeled \textit{d} in Fig.~\ref{Fig:Interfaces}). The surface energy difference between (2$\times$2)$_{\text{G}}$- and the (3$\times$3)$_{\text{G}}$-interface amounts to $0.13$~eV at the graphite line, favoring the (2$\times$2)$_{\text{G}}$-interface.

To put our results into context, we revisit the growth process. In experiment, growth starts with a clean (3$\times$3) reconstruction. The sample is annealed until the surface is covered by graphene. At this stage, the (3$\times$3) reconstruction is the dominant phase underneath graphene.\cite{hiebel2012sas} However, a shift from the SiC (3$\times$3) to the (2$\times$2)$_\text{C}$ surface reconstruction at the graphene/ SiC interface can be stimulated by an additional annealing step at a temperature below graphitisation (950C$^\circ$ - 1000C$^\circ$) leaving the graphene layer unaffected.\cite{hiebel2012sas} Graphene growth starts at a point where bulk SiC decomposes. The interface is determined by the momentary stoichiometry at which the sublimation stopped - a coexistence of different phases is observed. The final annealing step shifts the chemical potential into a regime in which SiC bulk decomposition stops and the graphene layer does not disintegrate, but the SiC surface at the interface moves closer to local equilibrium, in agreement with the phase diagram in Fig.~\ref{Fig:Interfaces}, forming an interface structure between the (2$\times$2)$_\text{C}$ reconstruction and graphene. 

In summary, we shed new light on the central aspects of the thermodynamically stable phases that govern the onset of graphene formation on SiC($\bar{1} \bar{1} \bar{1}$). In our view, the main difference between the Si face and the C face is the fact that Si-terminated phases are more stable on the C face due to the formation of heterogeneous Si-C bonds. The Si rich phases are thus stable practically up to the graphite line, allowing the surface to form graphene only at the point where bulk SiC itself is no longer stable. This makes monolayer graphene growth on the C side more difficult than on the Si face.

\begin{acknowledgments}
The authors thank Dr.~Fanny Hiebel, Dr.~Ulrich Starke, and Renaud Puybaret for fruitful discussions during the preparation of this work. This work was partially supported by the DFG collaborative research project 951 ``HIOS''.
\end{acknowledgments}

\end{document}